# TPIFM: A Task-Aware Model for Evaluating Perceptual Interaction Fluency in Remote AR Collaboration

Jiarun Song, *Member, IEEE*, Ninghao Wan, Fuzheng Yang, *Member, IEEE*, Weisi Lin, *Fellow, IEEE*

***Abstract*—Remote Collaborative Augmented Reality (RCAR) enables geographically distributed users to collaborate by integrating virtual and physical environments. However, its reliance on real-time transmission makes it susceptible to delay and stalling impairments under constrained network conditions. Perceptual interaction fluency (PIF), defined as the perceived pace and responsiveness of collaboration, is influenced not only by physical network impairments but also by intrinsic task characteristics. In particular, different tasks impose distinct temporal requirements and cognitive demands, which jointly determine a task-specific just-noticeable difference (JND), i.e., the maximal tolerable temporal responsiveness before PIF degrades. This JND serves as a perceptual threshold that relates network impairments to PIF. When the average response time (ART), measured as the mean time per operation from receiving collaborator feedback to initiating the next action, falls within the JND, PIF is generally sustained, whereas values exceeding it indicate disruption. From the perspective of the Free Energy Principle (FEP), tasks with lower JNDs impose stricter temporal prediction demands, making PIF more sensitive to impairments, whereas higher JNDs allow greater tolerance. On this basis, we classify RCAR tasks by JND and evaluate their PIF through controlled subjective experiments. Building on these findings, we propose the Task-Aware Perceptual Interaction Fluency Model (TPIFM) to evaluate PIF under network impairments in RCAR scenarios. Experimental results show that TPIFM significantly outperforms state-of-the-art baselines, achieving improvements of 6.0% and 13.6% in PCC and SROCC, respectively, along with a 51.6% reduction in RMSE. These results offer both theoretical insights and practical guidance for the design of adaptive RCAR system and the optimization of user experience.**

## I. Introduction

With advances in network technology, augmented reality (AR) now better supports real-time remote operation and collaboration [1]-[5]. Remote Collaborative AR (RCAR) enables geographically distributed users to jointly perform tasks by seamlessly integrating virtual content into physical spaces. Unlike 2D video-based collaboration, RCAR provides more intuitive interaction with real environments, enabling easier object manipulation and spatial coordination [6], [7]. Users can perform actions like hand-based operations and path planning in a natural and context-aware way [8]. Moreover, AR focusing on supporting tasks through visual feedback and a shared spatial context, which enhances decision-making and situational awareness [9]. As a result, RCAR is widely used in remote maintenance, collaborative design, training, medical guidance, and emerging application areas [10]-[12].

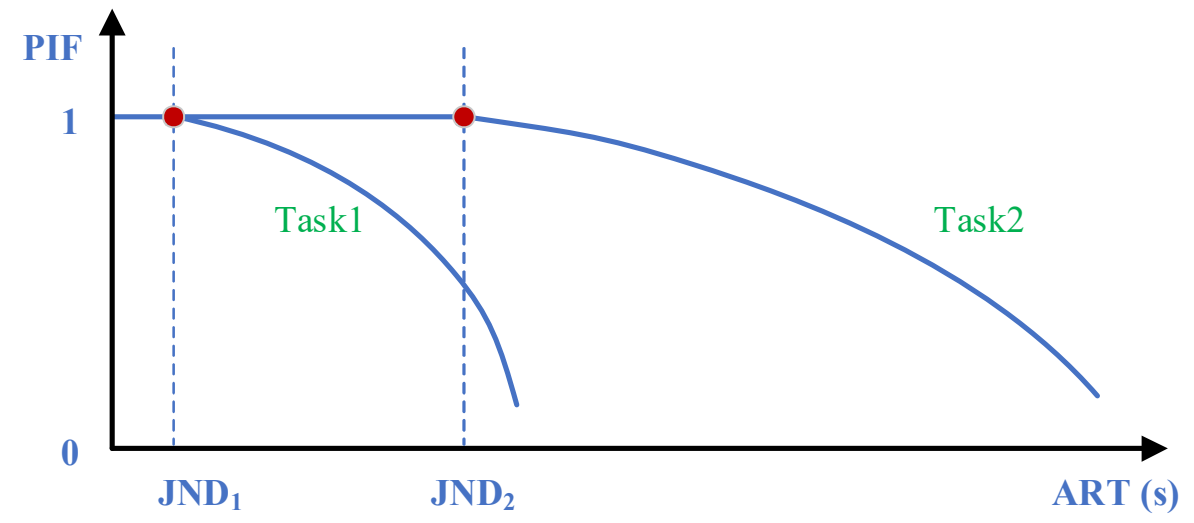

Fig. 1. Illustration of JNDs for two tasks.

RCAR systems primarily support real-time interpersonal collaboration, where the perceptual interaction fluency (PIF), defined as the subjectively perceived pace and responsiveness of collaboration, is a key dimension of Quality of Experience (QoE) [13], [14]. However, network impairments such as latency, jitter, and packet loss often introduce response delays and stalling effects [15]-[17], which directly degrade PIF [18]. Importantly, the impact of these impairments varies across tasks, as users' sensitivity and tolerance to temporal disruptions depend on the intrinsic task characteristics, such as temporal requirements and cognitive demands. These characteristics are reflected in a task-specific just-noticeable difference (JND) [19]-[22], defined as the maximal tolerable temporal responsiveness before PIF begins to degrade. The JND serves as a perceptual threshold that relates network impairments to PIF.

As illustrated in Fig. 1, we consider two representative task types. Task 1 is a high-tempo, low-cognitive-load turn-taking task that requires near-immediate responses, yielding shorter response times and a lower JND ($JND_1$). When the actual average response time (ART) exceeds $JND_1$, users are more likely to perceive a decline in PIF, making such tasks highly sensitive to impairments. In contrast, Task 2 is a deliberative, turn-based decision-making task that involves a higher cognitive load, leading to longer response times and a higher JND ($JND_2$), which provides greater tolerance to temporal deviations and a more resilient PIF. This difference can be explained from the perspective of the Free Energy Principle

This work was supported in part by the National Natural Science Foundation of China (62171353) *(Corresponding author: Weisi Lin).*

Jiarun Song and Ninghao Wan are with the School of Telecommunications Engineering, Xidian University, Xi'an, 710071, China (e-mail: jrsong@xidian.edu.cn; ninghaow@stu.xidian.edu.cn).

Fuzheng Yang is with the School of Telecommunications Engineering, Xidian University, China, and with the School of Electrical and Computer Engineering, Royal Melbourne Institute of Technology, Melbourne, VIC 3001, Australia (e-mail: fzhyang@mail.xidian.edu.cn).

Weisi Lin is with the College of Computing and Data Science, Nanyang Technological University, Singapore 639798 (e-mail: wslin@ntu.edu.sg).

(FEP) [23]-[24], which states that the brain minimizes prediction errors by continuously updating its internal model. In RCAR scenarios, network-induced delays and stalling impairments disrupt temporal predictions, thereby increasing prediction errors and prompting perceptual or behavioral adjustments. Tasks with lower JND impose stricter temporal-prediction requirements, leading to greater sensitivity to impairments, whereas tasks with higher JND allow greater tolerance and resilience. From a theoretical perspective, this behavior can be interpreted using the FEP, which provides a conceptual explanation for task-dependent differences in temporal sensitivity, rather than serving as an operational component of the model formulation. This observation is also consistent with prior findings that cognitive processes and task engagement play an important role in shaping collaborative performance in remote teamwork [26].

Despite these insights, existing studies still face several limitations. First, most works focus on network impairments themselves, without explicitly considering how intrinsic task characteristics influence user perception. Second, task-specific temporal tolerance, such as JND, is rarely incorporated into the evaluation of interaction fluency. Third, there is a lack of unified modeling approaches to quantify how multiple impairments (e.g., delay and stalling) jointly affect PIF under diverse task conditions. These limitations highlight the need for a task-aware modeling approach for PIF in RCAR systems.

To address this issue, we investigated how task-specific JND moderates the impact of delay and stalling on PIF in RCAR scenarios. Specifically, a series of controlled subjective experiments was conducted on representative RCAR tasks with different JND levels. The effects of delay, stalling, and their combination on PIF across tasks were systematically examined, with delay and stalling represented by end-to-end (E2E) latency and the stalling ratio, respectively. Based on these findings, a Task-Aware Perceptual Interaction Fluency Model (TPIFM) was proposed to evaluate PIF under network impairment in RCAR scenarios. The present study focuses on controlled single-task settings to establish task-dependent perceptual characteristics, providing a basis for future extension to more complex RCAR interaction scenarios. Fig. 2 illustrates the framework of the proposed model. The main contributions of this work are as follows:

(1) We introduce task-specific JND as an intrinsic attribute to characterize temporal responsiveness requirements across RCAR tasks, enabling a task-aware analysis of PIF.

(2) We conduct systematic subjective experiments to quantify how JND moderates the effects of delay, stalling, and their combination on PIF,

(3) We propose a task-aware evaluation model (TPIFM) that incorporates JND to accurately quantify PIF under diverse RCAR scenarios.

The rest of this paper is organized as follows: Section II reviews related work. Section III presents the experimental design. Section IV analyzes the results and introduces the proposed TPIFM model. Section V discusses the model's performance, and Section VI concludes the paper.

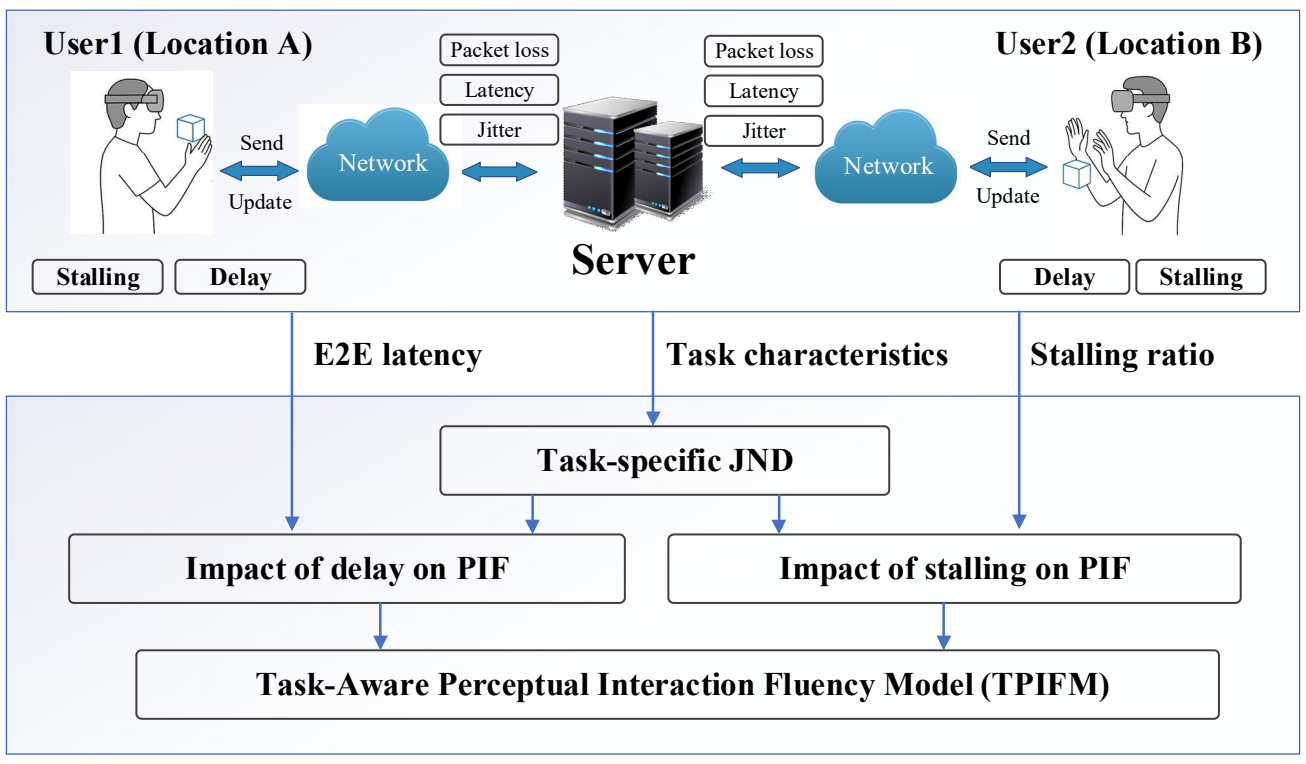

Fig. 2. Framework of perceptual interaction fluency evaluation model.

## II. Related Work

The following section provides a brief overview of the related work, focusing on the development of RCAR services and QoE evaluation in AR-based applications.

### *A. Development of RCAR Services*

In recent years, RCAR services have shown a diversified trend in both technological advancement and application expansion. RCAR has been widely adopted across key domains such as industrial maintenance, emergency management, medical collaboration, and educational innovation [27]-[29]. For example, in industrial maintenance, RCAR systems enable remote experts to provide real-time annotations on equipment repair procedures via AR glasses. Virtual instructions are spatially registered and overlaid on the physical surfaces of machinery, allowing on-site technicians to follow these step-by-step guides. This significantly improves operational efficiency and accuracy [4], [27]. In medical collaboration, RCAR technology enables real-time projection of 3D anatomical structures onto patients' bodies or surgical areas. Multiple doctors can simultaneously view and interact with shared virtual surgical models using AR devices, supporting remote surgical planning and collaborative procedures with enhanced precision [28]. In educational innovation, RCAR enables the overlay of 3D models (e.g., geological formations or molecular structures) onto laboratory tools or teaching environments. Using AR glasses, teachers and students can engage in immersive knowledge transfer and interactive learning experiences, such as jointly exploring terrain changes, thereby enhancing the intuitive understanding of abstract concepts [29].

To address the diverse characteristics of RCAR services, prior studies have proposed various classification frameworks and system designs. For example, Sereno et al. [2] introduced a multi-dimensional classification framework encompassing factors such as time, space, role symmetry, and interaction devices, providing a structured basis for analyzing collaborative AR experiences. Building on this, subsequent work has emphasized context-aware modeling and system adaptability [8], enabling dynamic adjustment of collaborative interactions in complex environments. These theoretical advancements have facilitated practical implementations, such as adaptive visual instruction systems [30] and collaborative XR platforms [31] that maintain real-time consistency. In addition, low-latency rendering and transmission techniques have been explored to support responsive interaction in remote collaboration scenarios [32]. While these developments provide a structural and technical foundation for RCAR, they also introduce increased complexity in ensuring a high-quality user experience across diverse tasks and system configurations. Understanding how these system and task factors influence perceived interaction quality motivates the need for dedicated QoE evaluation.

### *B. QoE Evaluation of AR-based Services*

For RCAR services, delay and stalling are among the most common impairment effects that negatively impact interaction. Although prior studies have tried to reduce the impact of network impairment as much as possible through improvements in interaction algorithms [33]-[35] and optimizations of distributed network architectures [36]-[41], these impairments remain difficult to eliminate entirely. Therefore, it is essential to investigate the impact of delay and stalling on QoE in multi-user RCAR scenarios.

Regarding latency-induced impairments, most existing studies have focused on single-user human-computer interactions, using objective metrics such as task completion time and error rate [42]-[46]. Recent work has expanded to include user perception, providing latency thresholds across varying contexts. For example, in first-person shooter games, noticeable impairment begins at 75 ms RTT, with severe degradation at 400 ms [47]. In cloud-based social XR systems, Mitra et al. [48] conducted extensive subjective tests involving factors such as RTT, jitter, and packet loss, finding that latency within 77 ms has minimal impact on QoE. In mobile AR, users notice latency differences at 69 ms, yet prefer systems below 132 ms, even though performance remains unaffected [49]. For XR tasks such as block-building over video conferencing, acceptable experience is maintained only if E2E latency remains under 900 ms [50]. Meanwhile, AR interaction thresholds vary with task demands, ranging from around 75 ms for online gaming to up to 250 ms for telemetry involving the vestibular system [51]. Despite these insights, current approaches remain largely task-specific and fragmented, with reported latency thresholds varying significantly across task types and interaction contexts and lacking unified, systematic evaluation standards. This highlights the need for a more structured latency-impact assessment framework that accounts for task urgency and perceptual sensitivity.

Regarding the impact of stalling, existing research primarily focuses on optimizing network performance and designing lightweight E2E AR system architectures [52]-[54]. However, there remains a lack of systematic investigation into how interaction stalling affects user experience quality in multi-user collaborative AR scenarios. For example, [55] highlights that the impact of stalling may vary depending on users' roles within a collaborative task, and that providing clear information about poor network conditions can help mitigate some of the negative effects. The study [56] further explores the differential impact of stalling under task-based and task-free conditions. It finds that users are more sensitive to stalling in non-task situations. This conclusion was not drawn from experiments involving users' active participation in interaction tasks, limiting its generalizability to real-time collaborative contexts.

These existing studies largely rely on qualitative methods, such as post-experience questionnaires, and do not offer predictive or computational models of how stalling impairs interaction. Krogfoss et al. [17] proposed a QoE evaluation framework for XR applications over mobile networks that considers the effects of both stalling and delay. This work emphasizes the mapping between Key Quality Indicators (KQIs) and network-level Key Performance Indicators (KPIs). However, this work does not elaborate on how to design subjective experiments for specific XR scenarios to obtain KPI degradation models. More recently, Lee et al. [57] developed quantitative QoE models to drive adaptive transmission schemes in cloud-based VR systems, demonstrating the potential of predictive models for real-time system optimization. Despite these advances, these studies mainly focus on system-level optimization and architectural design, while largely overlooking the role of intrinsic task characteristics in shaping user perception. More importantly, a unified task-aware model is still lacking to quantify how task-dependent temporal sensitivity influences PIF under varying network impairments.

## III. Design of Experiments

To fulfill the objectives of this work, a collaborative task experimental test platform was developed, and a series of subjective experiments were conducted using this platform. Users' subjective ratings of PIF were recorded for model training and verification. The detailed settings of the experiments are presented in the following section.

### *A. Test Platforms*

Although RCAR systems have seen preliminary applications in areas such as industrial manufacturing and remote healthcare, the high complexity of interactive tasks, the presence of numerous interfering variables, and the lack of precise control interfaces for delay and stalling impairments make it difficult to ensure that evaluations are conducted under consistent conditions, which limits the quantitative analysis of PIF. To address these challenges, this study develops a multi-user, real-time, interactive AR test platform to provide a controllable and repeatable experimental environment. The platform has the following functions: (1) It supports flexible configuration of transmission strategies and realizes dynamic control of E2E latency through frame rendering timing management, (2) It can simulate stalling effects in weak network conditions, and accurately control the frequency and duration of stalling in the interaction process, (3) It integrates an automated rating interface with multi-dimensional data collection capabilities, enabling the synchronous recording of subjective rating scores, user actions, as well as system parameters to support subsequent modeling of PIF. The overall architecture of the platform is shown in Fig 3(a).

Specifically, the virtual object content on the experimental platform is developed using the Unity3D engine [58] and the Mixed Reality Toolkit (MRTK) [59]. The application is built through the Universal Windows Platform (UWP) and deployed to the Microsoft HoloLens 2 devices through Wireless LAN (Wi-Fi). Virtual objects are presented across multiple terminal spaces via a visual sharing mechanism, with data transmitted via asynchronous Socket API over TCP [60]. Fig. 3(b) and 3(d) show the synchronous perspective of remote users moving virtual objects together. The platform allows users to interact with virtual interfaces via hand gestures, which greatly enhances interaction flexibility.

To facilitate analysis and quantitative modeling, the platform enables precise control of impairment parameters. This study is based on controlled experiments rather than simulation with tunable hyperparameters. Therefore, no conventional simulation hyperparameters are involved. Instead, the key experimental variables, including E2E latency levels and stalling configurations, serve as controllable factors that define the experimental conditions and support systematic analysis. By managing the rendering time of each frame, the system dynamically regulates the average E2E latency. The inherent system latency is calculated by comparing Network Time Protocol (NTP) timestamps between the capture time at the sender and the rendering time at the receiver and is approximately 15 ms on average across 50 independent interaction tests. To adjust interaction delay, programmable additional latency is added to the inherent latency, enabling fine-grained control of E2E latency. To simulate the stalling effects, the platform extends the rendering duration of specific frames to precisely control the onset and duration of stalling events. During interaction, the system monitors the object's pose changes in real time. When stalling conditions are triggered, user input is suspended and the physics engine is frozen, causing a visual freeze. Once the stalling duration ends, system components and physics behaviors automatically resume. This design enables highly controllable, reproducible simulation of stalling, providing experimental support for studying its impact on PIF.

Additionally, the platform integrates a subjective evaluation interface and automatic data logging. After completing a collaborative task, users are prompted to rate the perceived impact of delay and stalling on PIF, as shown in Fig. 3(c) and Fig. 3(e). All delay and stalling parameters, along with user ratings, are recorded to support the development of evaluation models in AR environments.

*B. Participants*

A total of 48 participants aged 21 to 26 years (M=23.5, SD=1.2) were recruited voluntarily for this study, including 20 females (41.6%) and 28 males (58.4%). All were university students and had normal or corrected-to-normal vision and hearing. Six participants (12.5%) reported prior experience with AR devices, while the remaining 42 (87.5%) had none. The study did not collect personal or sensitive data, nor did it include invasive procedures. The tasks were limited to AR collaboration under controlled laboratory conditions and thus posed minimal risk to the participant. To minimize potential visual fatigue and physical discomfort associated with head-mounted displays, continuous testing in each experiment was restricted to 20 minutes per participant, after which a short break was provided before resuming [61].

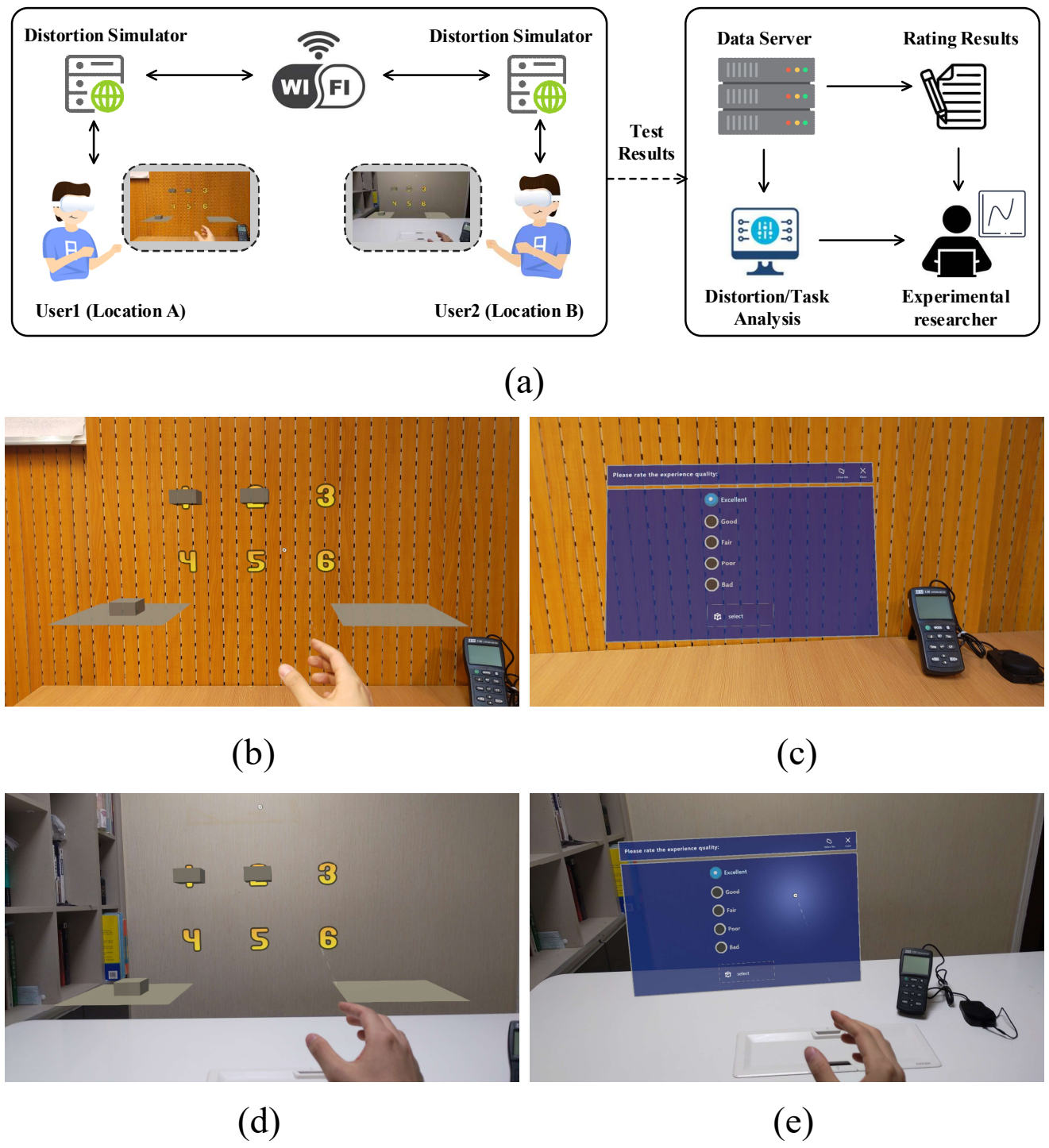

Fig. 3. Illustration of the designed test platform, (a) Architecture of the test platform, (b) User1's perspective, (c) User1's rating interface, (d) User2's perspective, (e) User2's rating interface.

*C. Experimental Settings*

The experiment consists of three parts. Experiment 1 focuses on task design and JND-based representation. Experiment 2 is designed for model training, investigating the effects of delay, stalling, and their combined impact on PIF. Experiment 3 is designed to validate the performance of the proposed model. Together, these experiments cover task characterization, model training, and validation.

*Experiment 1: Task Design and Representation*

Given that task-specific JND may affect the perception of impairments, it is essential to characterize and quantify this sensitivity to ensure accurate evaluation across tasks. To this end, six collaborative tasks with varying temporal demands were designed, as illustrated in Fig. 4. These tasks cover a broad spectrum of interaction patterns, including fast-paced, moderately paced, and deliberative interaction patterns to enable comprehensive analysis of task-dependent sensitivity to impairments. The details of each task type are described below:

(1) **Sudoku Puzzle (SP):** Two participants alternately collaborate to complete a 4×4 Sudoku puzzle. Each participant must observe the pattern and spend time thinking before operating, and can only operate after the other participant has completed their turn. This task emphasizes cognitive reasoning rather than rapid action, corresponding to a relatively high JND.

(2) **Tic-Tac-Toe (TTT):** Two participants alternately place their pieces on a 3×3 grid, aiming to align three pieces in a row, either horizontally, vertically, or diagonally. The first to complete such a line wins. This task requires participants to observe the opponent's move, consider strategies, and decide before placing the next piece. Although the thinking time is short compared with that for more complex tasks, it still involves reasoning and anticipating the collaborator's actions, resulting in slightly slower responses than reflexive tasks.

(3) **Modular Arithmetic Relay (MAR):** This task is a collaborative activity based on modular arithmetic rules. In each round, a participant receives a random number and performs a modulo operation with a preset divisor. The result determines the block's placement (e.g., if 25 mod 6 equals 1, the block is placed in the first position). The task requires participants to calculate quickly and place blocks accurately, with the team aiming to complete an ordered arrangement under specific constraints. By adjusting parameters such as the divisor and the number range, the difficulty level can be flexibly adjusted to improve both computational efficiency and coordination logic.

(4) **Block Relay (BR):** This collaborative task involves rapid sequential actions, where participants alternately place 9 blocks in designated positions as quickly as possible. In each round, the system provides a clear target position instruction. Unlike the modular arithmetic domino task, this activity requires no calculations. Participants place the next block immediately after the other completes theirs, continuing until the task is finished. The task emphasizes operational continuity and pace coordination.

(5) **Laboratory Equipment Sorting (LES):** In this task, two participants collaboratively sort and organize laboratory equipment. One participant is responsible for identifying and selecting the equipment, while the other handles its placement. They take turns operating, and each step must be completed under the partner's supervision. The next action can begin only after the previous one is finished. This task emphasizes orderly coordination, mutual monitoring, and accurate execution.

(6) **Vehicle Assembly (VA)**: In this task, two participants work together to assemble a virtual vehicle by following a predefined sequence. Each participant takes turns installing different components in the correct positions. A participant must wait until their partner completes the current assembly step before proceeding. This task highlights the importance of stepwise execution and precision in collaborative operations.

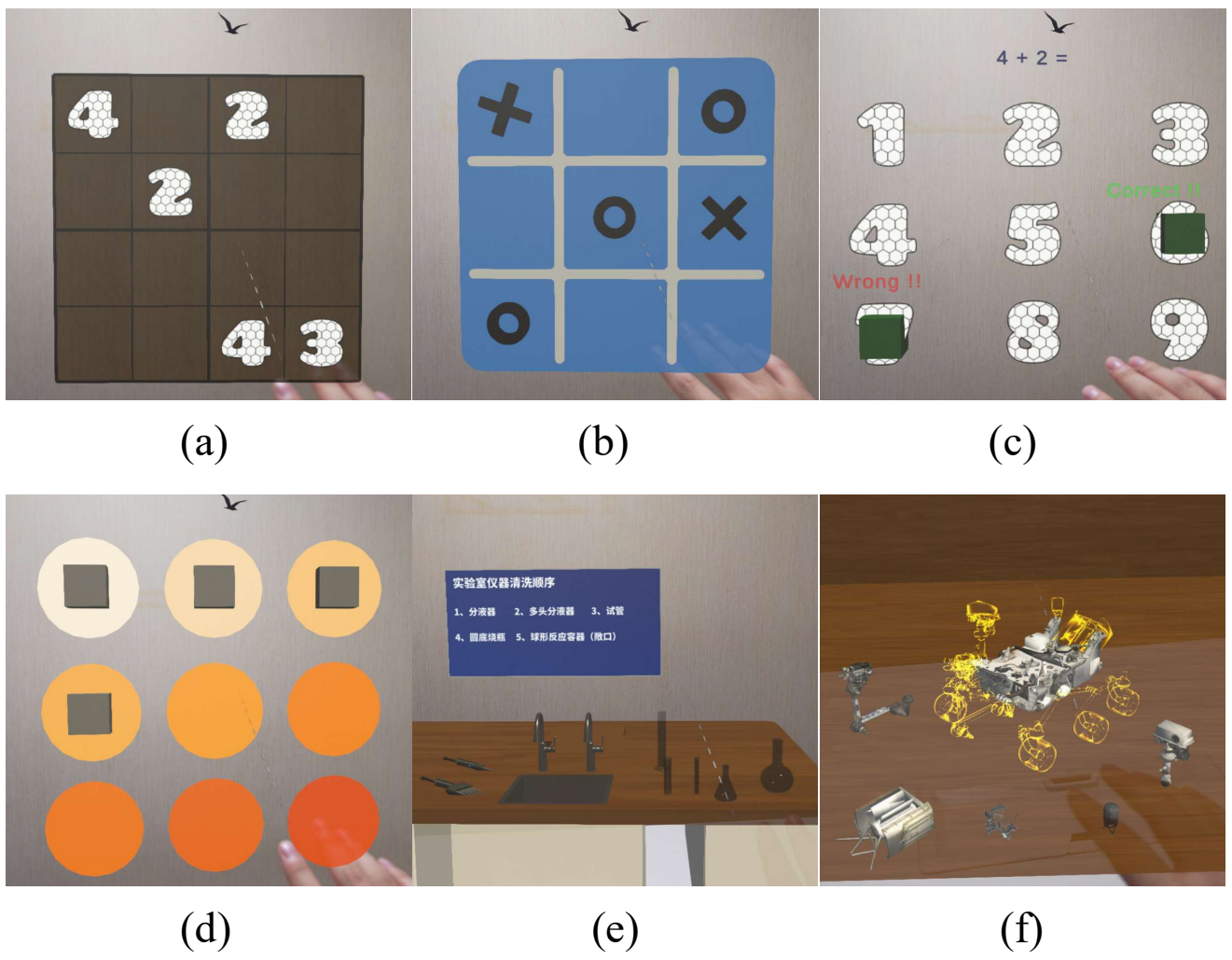

Fig. 4. Task illustration. (a) Sudoku Puzzle (SP), (b) Tic-Tac-Toe (TTT), (c) Modular Arithmetic Relay (MAR), (d) Block Relay (BR), (e) Laboratory Equipment Sorting (LES), (f) Vehicle Assembly (VA).

TABLE I
JND OF EACH TASK

| **Task** | SP | TTT | MAR | BR | LES | VA |
|---|---|---|---|---|---|---|
| **JND (s)** | 3.34 | 1.57 | 1.16 | 0.38 | 0.71 | 0.95 |

Although designed under controlled experimental settings, the selected tasks were intended to represent different interaction characteristics relevant to RCAR applications. The laboratory-style tasks (SP, TTT, MAR, and BR) capture different levels of cognitive processing and operational coordination. Specifically, SP, TTT, and MAR require participants to perform task-dependent reasoning or calculation before responding, whereas BR emphasizes rapid sequential operations with minimal cognitive delay. In contrast, the application-oriented tasks (LES and VA) extend the evaluation to more practical RCAR scenarios involving collaborative equipment operation and virtual assembly. To validate the rationality of interactive task settings, a dedicated experiment was conducted with 24 participants. The experiment had two objectives: (1) to assess the effectiveness of task design in differentiating tasks with varying temporal demands, and (2) to measure the JND for each task. These measurements offered empirical support for task classification.

A paired design was used, with participants completing tasks in pairs. The experimental process includes the following operations. In the pilot phase, participants were thoroughly briefed on the experimental procedure and task goals. Each pair was guided into separate experimental environments and assisted by professionals to complete the wearing and calibration of the HoloLens 2 headset. For those unfamiliar with AR devices, A five-minute period of free exploration was conducted to help them become acquainted with the device's operation. After that, participants were given detailed task instructions to ensure smooth execution. In the formal experiment, each pair completed three rounds of all six tasks under ideal network conditions (no delay or stalling). For each task, we computed the baseline average response time ($ART_j$) under ideal network conditions and used it as an operational approximation of the task-specific JND. This is because the baseline ART reflects the inherent temporal responsiveness required by each task. When additional impairments (e.g., delay) increase the actual ART beyond $ART_j$, the deviation from the task's expected responsiveness becomes noticeable, leading to PIF degradation. This is consistent with the role of JND as a perceptual threshold.

Table I reports the JND values for each task. A repeated-measures ANOVA showed significant differences across tasks ($F = 547.48$, $p < .001$, $\eta^2 = 0.996$), indicating that the task set effectively captures behavioral variation in temporal demand. These results provide a statistically validated basis for distinguishing JND levels across tasks and support the subsequent subjective experiments.

*Experiment 2*: *Data Collection for Model Training*

This experiment investigates how task-specific JND moderates the impact of delay and stalling on PIF in RCAR scenarios and aims to establish an objective evaluation model. Four task scenarios were tested, including SP, TTT, MAR, and BR. To minimize uncontrolled network interference, all tests were run over a LAN with stable transmission performance. Using reserved interfaces in the test platform, we introduced adjustable delay and stalling via controllable buffering, enabling reliable, level-specific impairment simulation. This setup facilitated a systematic, quantitative analysis of how different network impairments affect PIF, thereby providing empirical evidence to support the development of an objective evaluation model for PIF under diverse network conditions. Since participants in Experiment 1 had not experienced the testing scenarios of Experiment 2, the same participants were recruited for this study.

*Session 1*: *Experiment on the Impact of Delay*

This session uses E2E latency as the research variable for the delay effect and designs an experiment across different task scenarios. The experiment sets eight E2E latency levels: 100 ms, 200 ms, 300 ms, 500 ms, 800 ms, 1,000 ms, 2,000 ms, and 3,000 ms. The selected latency levels cover a wide range of perceptual conditions, from near-imperceptible delay to severe impairment, enabling comprehensive analysis of how increasing latency degrades PIF. To control for order effects, a Latin square design [62] was adopted to randomize latency levels across multiple groups, thereby improving the reliability and validity of the results.

The experiment consisted of two stages: a pilot study and the formal experiment. In the pilot phase, a stepped-delay exposure design was adopted, using the BR task as an example. Three latency levels (100 ms, 800 ms, and 3000 ms) were presented in sequence to establish a baseline for PIF. These levels were selected to represent low, moderate, and severe delay conditions, allowing participants to form a clear perceptual reference across different impairment levels. A five-point Likert scale was constructed in accordance with the ITU-T P.910 Absolute Category Rating (ACR) method [61] to evaluate PIF. The scale consisted of the following levels: "Perfect", "Good", "Fair", "Poor", and "Bad". After each test sequence, participants rated the PIF. The pilot study served to familiarize participants with the task and to help them form an initial understanding of the types of impairment involved.

In the formal experiment, participants completed all tasks in a randomized order. Within each task, they randomly experienced 8 different delay test scenarios. After each test scenario, participants were required to subjectively evaluate the PIF affected by delay. When all the scenarios are evaluated, the current task ends. Participants then removed the HoloLens 2 headset and took a five-minute break before proceeding to the next task. This process was repeated until all tasks were completed, marking the end of the experiment. A total of 768 subjective ratings (8 scenarios × 4 tasks × 24 participants) were collected. Based on these ratings, the Mean Opinion Score (MOS) [60] was calculated for each testing condition by averaging the corresponding subjective scores across participants, resulting in 32 MOS values for analysis.

*Session 2*: *Experiment on Impact of Stalling*

In this experiment, the effects of stalling on PIF were investigated by manipulating two variables, average stalling duration and stalling frequency, to produce varying levels of impairment across different tasks. The average stalling duration is the mean length of each stalling event, while the stalling frequency is the number of such events per interaction. To ensure the perceptual salience of the impairment, stalling events were triggered within the interaction cycle, specifically 500 milliseconds after the user issued an input command. Each interaction cycle was limited to a single stalling event to avoid overlapping.

The experimental parameters were defined as follows: the average stalling duration in each task was set to six levels: 100 ms, 200 ms, 300 ms, 500 ms, 1,000 ms, and 2,000 ms. The stalling frequency was set at 1, 2, or 3 events per task. These parameter values were selected to cover a wide range of stalling severity, enabling systematic analysis of their perceptual impact on PIF. Using a Latin square design, 14 representative parameter combinations were selected to cover a full range of test conditions, from instantaneous mild stalling to long-lasting, severe stalling. The specific stalling parameters for each task are presented in Table II.

Similar to the delay experiment, this test consisted of two phases: a pilot study and a formal experiment. In the pilot phase, a stepped stalling exposure design was employed using the BR task as an example. Three levels of stalling parameters representing perfect to poor interaction conditions were presented to establish a baseline PIF. In the formal experiment, participants completed all tasks in a randomized order. Within each task, they experienced 14 different interaction scenarios, each corresponding to a unique stalling configuration. After completing each scenario, participants provided a subjective rating of PIF based on their perceived stalling effect. Once all scenarios in a task were evaluated, the task was considered complete. Participant then took a five-minute break before the next task. This process was repeated until all tasks were completed. A total of 1,344 subjective ratings (14 scenarios × 4 tasks × 24 participants) were collected. Following the same data processing procedure as in Session 1, 56 MOS values were calculated.

TABLE II
PARAMETERS SETTINGS OF STALLING IMPAIRMENT

| Index | Times | Duration(ms) | Index | Times | Duration(ms) |
|---|---|---|---|---|---|
| 1 | 1 | 100 | 8 | 2 | 200 |
| 2 | 1 | 200 | 9 | 2 | 500 |
| 3 | 1 | 300 | 10 | 2 | 1000 |
| 4 | 1 | 500 | 11 | 3 | 100 |
| 5 | 1 | 1000 | 12 | 3 | 200 |
| 6 | 1 | 2000 | 13 | 3 | 500 |
| 7 | 2 | 100 | 14 | 3 | 1000 |

TABLE III
PARAMETERS SETTINGS OF DELAY AND STALLING IMPAIRMENT

| Index | Delay (ms) | Times of stalling | Avg duration (ms) |
|---|---|---|---|
| 1 | 1000 | 1 | 300 |
| 2 | 500 | 1 | 1000 |
| 3 | 2000 | 1 | 2000 |
| 4 | 800 | 2 | 500 |
| 5 | 1000 | 2 | 1000 |
| 6 | 2000 | 2 | 1000 |
| 7 | 300 | 3 | 500 |
| 8 | 2000 | 3 | 1000 |

TABLE IV
CONDITION SETTINGS FOR VALIDATION

| Index | Delay (ms) | Times of stalling | Avg duration (ms) |
|---|---|---|---|
| 1 | 200 | 1 | 200 |
| 2 | 2000 | 1 | 500 |
| 3 | 800 | 1 | 1000 |
| 4 | 500 | 1 | 2000 |
| 5 | 800 | 2 | 300 |
| 6 | 1000 | 2 | 400 |
| 7 | 400 | 2 | 800 |
| 8 | 200 | 3 | 200 |
| 9 | 600 | 3 | 400 |
| 10 | 1000 | 3 | 1000 |

*Session 3: Experiment on Impact of Combination Effects*

This experiment aimed to investigate the impact of combined delay and stalling impairments on PIF in RCAR scenarios. Participants were asked to complete the same set of collaborative tasks as in the previous experiments, where both types of impairment occurred concurrently. The delay and stalling parameters were defined as shown in Table III. For each task, participants experienced 8 test scenarios, each with a unique combination of delay and stalling levels.

The experimental procedure was consistent with that of the previous sessions. During the pilot phase, participants experienced three example trials to familiarize themselves with the experiment procedures. In the formal test, participants completed 8 randomized scenarios for each task. After each scenario, they rated PIF values using a five-point Likert scale, capturing the combined effects of delay and stalling. A total of

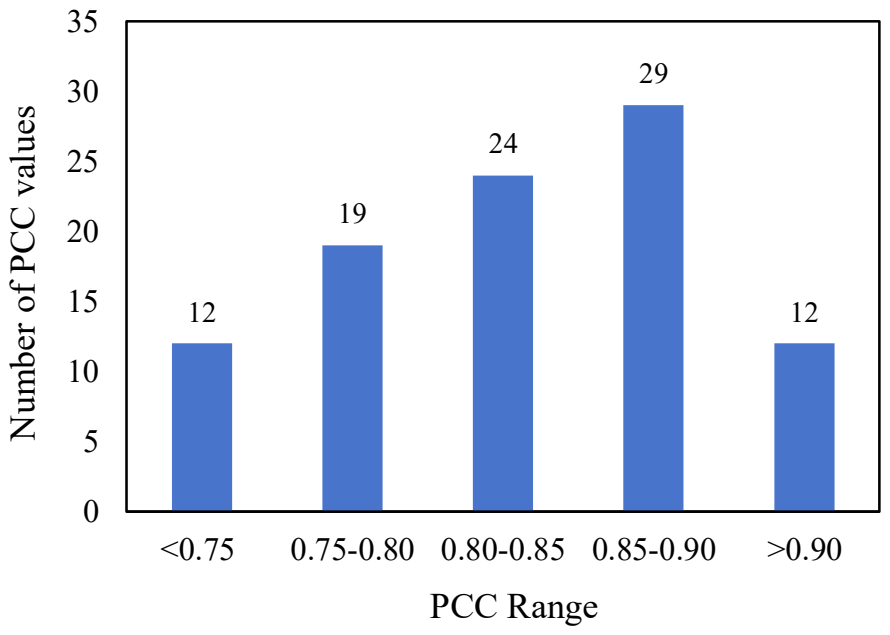

Fig. 5. Distribution of PCC values.

768 subjective ratings (8 scenarios × 4 tasks × 24 participants) were collected. Following the same procedure, 32 MOS values were calculated for subsequent analysis.

*Experiment 3: Data Collection for Validation*

To evaluate the proposed evaluation model, an additional test was conducted with 24 new participants who had not participated in any of the previous experiments. In addition to the original TTT and BR tasks, two application-oriented tasks were introduced: a LES task and a VA task, both designed to better reflect realistic RCAR scenarios. All test scenarios included both delay and stalling impairments, with parameter settings different from those used in Session 3 of Experiment 2. The specific configurations are listed in Table IV. These configurations were designed to represent different types and levels of impairment, enabling systematic evaluation of their combined effects on PIF and validating the applicability of the proposed model across diverse conditions. The experimental procedure followed the same protocol as previous sessions. A total of 960 subjective ratings (10 scenarios × 4 tasks × 24 participants) were collected. Following the same data processing procedure, 40 MOS values were calculated by averaging participant ratings for each testing condition and were used for model validation.

## *D. Subjective Data Reliability Analysis*

To assess the reliability of the collected subjective ratings, Pearson Correlation Coefficients (PCCs) were calculated between each participant's ratings and the corresponding MOS following ITU-T P.910 [61]. For experiments involving multiple task scenarios, ratings collected across all task scenarios were combined to obtain an overall PCC for each participant. A PCC threshold of 0.75 was adopted to identify participants for further inspection [61].

Fig. 5 illustrates the distribution of PCC values obtained from the training and validation experiments. Among the 96 PCC values, 84 (87.5%) exceeded the threshold of 0.75, while 65 (67.7%) were above 0.80, indicating generally good consistency between individual ratings and the corresponding MOS values. Although 12 PCC values fell below the threshold, further examination revealed no abnormal rating patterns. Therefore, all participants were retained for subsequent analysis. These results demonstrate that the collected subjective data provide a reliable basis for subsequent model training and validation.

## IV. Experimental Results for Modeling

This section first analyzes how task-specific JND moderates the impact of delay and stalling impairments on PIF in RCAR scenarios. Based on these observations, we propose TPIFM, whose formulation is derived from and fitted to the collected experimental data to effectively characterize PIF.

### *A. Task-based Modulation of Delay Impacts*

Before analyzing the impact of delay on PIF across tasks, the Shapiro-Wilk test [63] was conducted to verify the normality of participants' scores. The results confirmed that all four tasks satisfied the normality assumption (SP: $p = 0.255$, TTT: $p = 0.308$, MAR: $p = 0.446$, BR: $p = 0.610$; all $p > 0.05$), supporting the use of parametric analysis. A repeated-measures ANOVA was then performed, revealing significant differences in MOS across tasks ($p < 0.05$), indicating that the effect of delay on PIF is task-dependent.

Fig. 6 illustrates the trend of PIF (MOS) with increasing delay (E2E latency) across tasks. In the SP scenario, MOS remained above 3.8 as long as latency did not exceed 3s, indicating minimal impact on PIF. For TTT, MOS remained above 4 when latency was within 1,000 ms and above 3 for delays shorter than 3,000 ms. Overall, as latency increased from 100 ms to 3,000 ms, the MOS decreased gradually, in a nonlinear manner, from 4.6 to 3, reflecting a relatively modest drop of approximately 1.6 points. For the MAR task, MOS maintained above 4 when latency was below 450 ms, and the 3-point threshold was reached at around 1,200 ms. Over the full latency range (100 ms to 3,000 ms), the MOS dropped significantly from 4.7 to 1.8, with a total decline of nearly 3 points. BR task showed the highest sensitivity to latency. MOS stayed above 4 when latency was below 400 ms, while the 3-point threshold occurred between 600 and 800 ms. When latency exceeded 2,000 ms, the MOS fell to 2. The MOS plummeted from 4.7 to 1.4 as latency increased from 100 ms to 3,000 ms, marking a drop of over 3 points. Notably, the steepest decline occurred within the 100-800 ms range, after which the degradation trend became more gradual.

The differences in delay sensitivity across tasks are partially attributable to how users cognitively perceive and interpret delays. For example, in the TTT task, participants may introduce substantial subjective delays between interactions due to ongoing cognitive activities such as predicting the opponent's intentions and formulating a move strategy. These delays are inherently non-deterministic, as participants cannot precisely anticipate the opponent's response timing after making a move. This uncertainty leads to greater variability in response timing and raises users' tolerance thresholds, thereby reducing their sensitivity to system-induced delay variations, making network-induced delays less perceptually disruptive to PIF. In contrast, the BR task requires participants to take turns as quickly as possible, with each participant initiating their move immediately after the partner completes their actions. The absence of deliberate cognitive delay in this task led to a shorter average response time. Consequently, a heightened sensitivity to even minor delay impairments.

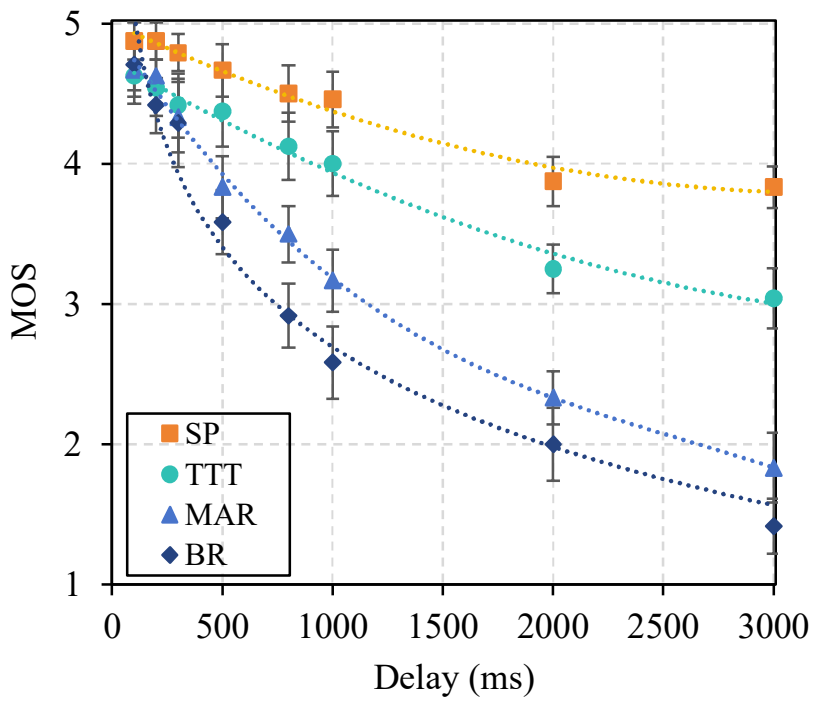

Fig. 6 Relationship between delay and PIF.

TABLE V
Parameters of $v_1$ and $v_2$

| Task | JND(s) | $v_1$ | $v_2$ | *R-square* | *RMSE* |
|---|---|---|---|---|---|
| SP | 3.34 | 4.786 | 7.99e-05 | 0.892 | 0.142 |
| TTT | 1.57 | 4.678 | 1.56e-04 | 0.981 | 0.089 |
| MAR | 1.16 | 4.777 | 3.57e-04 | 0.982 | 0.153 |
| BR | 0.38 | 4.764 | 4.86e-04 | 0.956 | 0.273 |

Based on the subjective data in Fig. 6, the correlation between PIF and delay was analyzed, showing a negative exponential decay trend with increasing delay. Accordingly, an objective evaluation model was established via least-squares fitting, as expressed below:

$$Q_d = \max\left(\min\left(v_1 \cdot \exp(-v_2 \cdot T_d),\ 5\right),\ 1\right) \tag{1}$$

where $Q_d$ represents the PIF affected by delay, $T_d$ denotes the E2E latency. $v_1$ and $v_2$ are fitted parameters. The parameters for each task are listed in Table V.

As shown in Table V, parameter $v_1$ remains approximately constant across different JND levels. Therefore, $v_1$ can be approximately equal to the mean of each task, which is 4.751. In contrast, $v_2$ gradually decreases as JND increases. The relationships between the JNDs and parameters $v_2$ can be expressed as follows:

$$v_2 = \alpha \cdot \exp\left(-\beta \cdot JND\right) \tag{2}$$

where $\alpha$ = 6.43e-04 and $\beta = 0.679$ controls the variation trend of $v_2$ with respect to JND. These parameters are obtained by least-squares fitting of the data in Table V.

### *B. Task-based Modulation of Stalling Impacts*

Before analyzing the impact of stalling impairment on PIF across tasks, a Shapiro-Wilk test was conducted to assess whether the participant's scores in each task followed a normal distribution. The results confirmed normality for all tasks (all p-values $> 0.05$). Based on this, a repeated-measures ANOVA was performed to examine whether significant differences existed between tasks. The analysis revealed statistically significant variations in MOS scores across tasks ($p < 0.05$), indicating that the impact of stalling on PIF differed significantly depending on the task scenarios.

To facilitate analysis of stalling impairment, this study uses the stalling ratio $R_s$ to represent different levels of stalling impairment. This ratio is defined as:

$$R_s = \frac{T_s}{T_s + T_m} = \frac{n \cdot T_a}{n \cdot T_a + T_m} \quad (3)$$

where $T_s$ denotes the total duration of stalling events during the task, $T_m$ represents the duration of user interaction actions, $T_a$ indicates the average duration of a single stalling event, and $n$ is the number of stalling events. These variables involved in $R_s$ are directly measured by the experimental platform.

Fig. 7 presents the PIF for each task under different stalling ratios. The results indicate that, at the same stalling level (stalling ratio), the decrease in MOS is relatively small in the SP task, suggesting greater tolerance to stalling impairment. In contrast, in the TTT and MAR tasks, the MOS drops more noticeably, reflecting a stronger impact of stalling on PIF. The BR task shows the steepest decline, with MOS values decreasing rapidly as $R_s$ increases, indicating that PIF is highly sensitive to stalling in this task. This result aligns with expectations, indicating that task-specific JNDs play a key role in shaping PIF. In tasks with low JNDs, such as BR, interactions occur in rapid succession, and participants often form an expected pace. In such cases, a stalling event can interrupt this pace and affect subsequent actions, amplifying the perceived disruption. By contrast, in tasks with high JNDs, such as SP and TTT, longer intervals between interactions allow more time for cognitive processing, reducing users' sensitivity to brief disruptions and thereby lessening the impact of stalling on PIF.

Based on the subjective evaluation results from the stalling experiment, the PIF decreases with $R_s$ following a negative exponential trend. Using the least squares method, the relationship can be expressed as follows:

$$Q_s = \max\left(\min\left(v_3 \cdot \exp(-v_4 \cdot R_s),\ 5\right),\ 1\right) \quad (4)$$

where $Q_s$ represents the PIF under stalling conditions, while $R_s$ denotes the stalling ratio. Parameters $v_3$ and $v_4$ are parameters obtained by least-squares fitting using the data in Fig. 7, as shown in Table VI. The parameter $v_3$ corresponds to the maximum MOS value. It can be seen that the maximum values across tasks are relatively close, and the $v_3$ value remains relatively stable, close to a constant of about 4.929. In contrast, $v_4$ is related to the curvature of the exponential function, and its variation across tasks reflects differences in the sensitivity of PIF to stalling impairments

As shown in Table VI, parameter $v_4$ gradually decreases as the JND increases. The relationship between $v_4$ and the JND can be expressed as follows:

$$v_4 = \rho \cdot JND^{-\sigma} \quad (5)$$

where, $\rho$ = 6.608, and $\sigma$ = 0.361 mainly control the variation trend of $v_4$ with JND. These parameters are obtained by least-squares fitting of the data in Table VI.

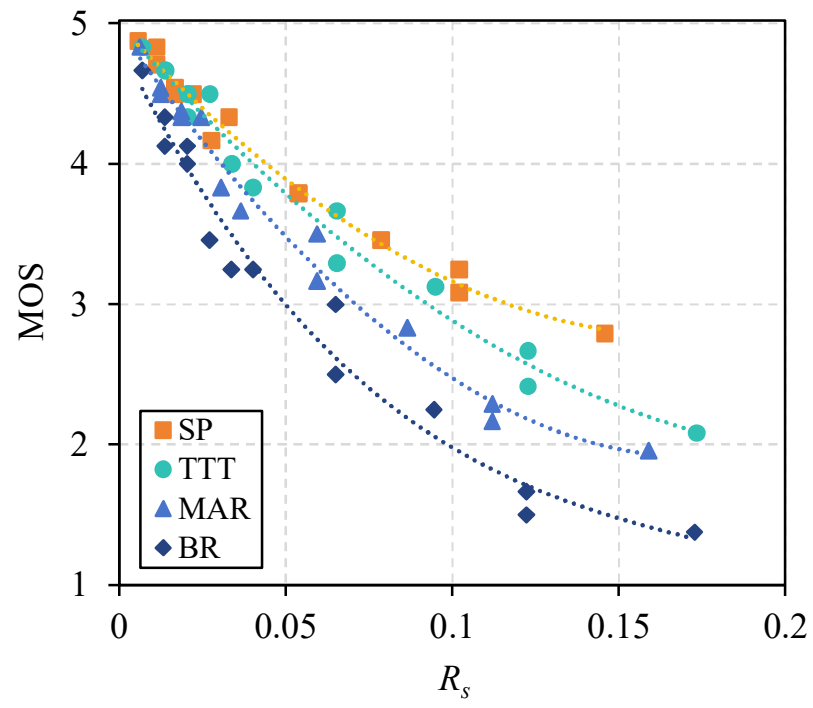

Fig. 7. Relationship between $R_s$ and PIF.

TABLE VI
PARAMETERS OF $v_3$ AND $v_4$

| Task | JND(s) | $v_3$ | $v_4$ | *R-square* | *RMSE* |
|---|---|---|---|---|---|
| SP | 3.34 | 4.905 | 4.282 | 0.976 | 0.110 |
| TTT | 1.57 | 4.965 | 5.301 | 0.976 | 0.145 |
| MAR | 1.16 | 4.911 | 6.611 | 0.978 | 0.147 |
| BR | 0.38 | 4.937 | 9.291 | 0.956 | 0.230 |

### *C. PIF evaluation model*

In this section, a PIF evaluation model was proposed for different RCAR tasks combining the impact of both delay and stalling effects. Considering both delay and stalling have negative impacts on PIF, the PIF under the combined influence of delay and stalling can be expressed as follows:

$$Q = \max\left(\min\left(4 \times (1 - v_5 \cdot (5 - Q_d) - v_6 \cdot (5 - Q_s)) + 1,\ 5\right),\ 1\right) \quad (6)$$

where $Q$ is the PIF affected by delay and stalling. $v_5$ = 0.104 and $v_6$ = 0.192 are weight values obtained by least-squares fitting using the data from Session 3 of Experiment 2. According to the fitting results, stalling appears to have a slightly stronger impact on PIF than delay. This may be because stalling is more likely to be perceived as a temporary interruption of the interaction process, whereas delay is generally perceived as a continuous slowdown of interaction, making stalling more disruptive to PIF.

In summary, this study focuses on RCAR applications and incorporates task-dependent characteristics to analyze the effects of delay, stalling, and their combined impact on PIF across different tasks. Based on these analyses, a comprehensive PIF evaluation model was developed for RCAR scenarios involving both delay and stalling. Moreover, the proposed model is computationally lightweight, as it involves only a fixed number of elementary operations. With task-specific JND estimated offline, it can be efficiently applied for online PIF evaluation.

## V. PERFORMANCE EVALUATION

To comprehensively evaluate the proposed model, we compare it with baseline models and perform ablation studies. The former evaluates the overall performance, while the latter examines the contribution of each component.

### *A. Comparison with Baseline Models*

Three baseline models that do not account for task-specific characteristics are considered for comparison (denoted as Baseline1, Baseline2, and Baseline3). Baseline 1 model is a linear model, Baseline 2 and Baseline 3 are non-linear model. Specifically, Baseline 1 adopts a linear formulation referenced from [16], which is defined as:

$$Q_{bl1} = \max\left(\min\left(a_1 + b_1 \cdot T + c_1 \cdot SR,\ 5\right), 1\right) \tag{7}$$

where parameters $b_1$ and $c_1$ indicate the weight of the delay and stalling effects, and $a_1$ is a constant term.

Baseline 2 follows the dual-exponential form proposed in [17], which is defined as:

$$Q_{bl2} = \max\left(\min\left(4 \times a_2 \cdot \exp(b_2 \cdot T) \times \exp(-\frac{SR}{c_2}) + 1,\ 5\right),\ 1\right) \tag{8}$$

where $T$ denotes the delay, and $SR$ represents the stalling ratio. The parameters $a_2$, $b_2$, and $c_2$ control the contributions of delay and stalling in the model.

Baseline 3 follows the ITU-T G.1072 modeling framework [66], which provides a general QoE aggregation structure for combining multiple impairment factors into a unified quality score. The delay-related impairment is modeled based on the formulation in [67], which defines a sigmoid-type function for delay-induced QoE degradation. The stalling-related impairment is modeled following [16], which characterizes temporal interruptions and their impact on perceived quality in interactive systems. The formulation is given as follows:

$$Q_{bl3} = \max\left(\min\left(5 - \frac{a_3}{1 + \exp(b_3 - c_3 \cdot SR)} - \frac{d_3}{1 + \exp(e_3 - f_3 \cdot T)},\ 5\right), 1\right) \tag{9}$$

Notably, all baseline model parameters were re-fitted using our experimental dataset via least-squares fitting. This ensures that each baseline is evaluated under consistent conditions and provides a fair basis for performance comparison. The fitted parameters are: $a_1$ = 4.21, $b_1$ = -0.0003, $c_2$ = -12.39 for Baseline1, $a_2$ = 1.07, $b_2$ = -0.24, $c_2$ = 0.15 for Baseline2, and $a_3$ = 2.48, $b_3$ = 2.29, $c_3$ = 38.71, $d_3$ = 2.48, $e_3$ = 3.19, $f_3$ = 0.0031 for Baseline3.

Three typical metrics specified by VQEG, namely the PCC, the Spearman Rank Order Correlation Coefficient (SROCC), and the Root Mean Squared Error (RMSE), were employed in the performance validation [64]. Generally, a smaller value of the RMSE and a larger value of the PCC and SROCC indicate superior performance. Table VII lists evaluation performance for these metrics across different validation datasets. The proposed model achieves higher PCC and SROCC values and lower RMSE than the baseline models, demonstrating its superior ability to predict PIF across different task scenarios. Furthermore, we followed [64] and performed an F-test on the residuals between the MOS and the predicted PIF to assess the statistical significance of the performance differences between the proposed and baseline models. The results show that all p-values are below 0.05, indicating that the observed improvements of the proposed model over the baseline models are statistically significant.

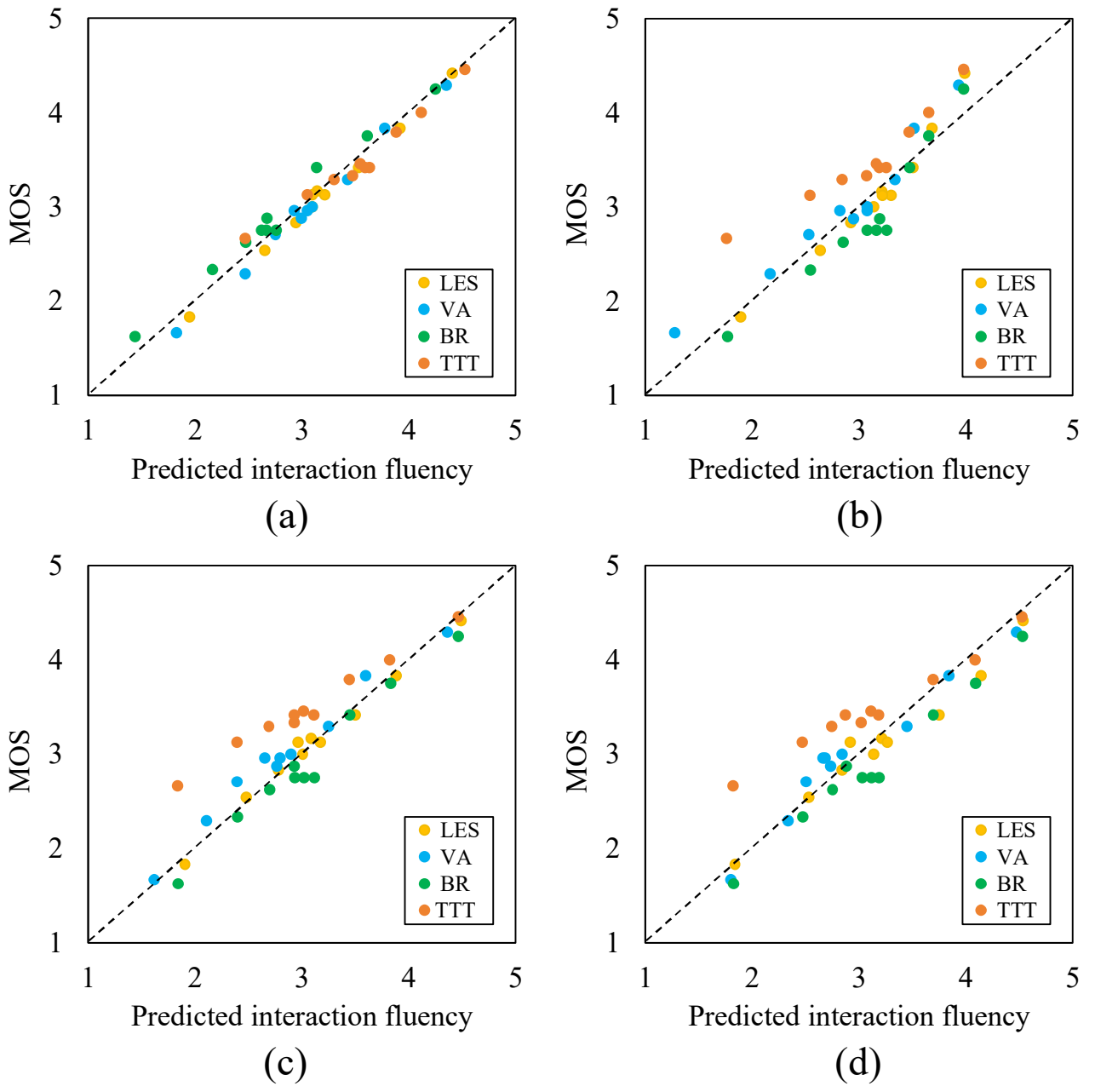

Fig. 8. Scatter plots of predicted PIF and MOS. (a) Proposed model, (b) Baseline1, (c) Baseline2, (d) Baseline3.

TABLE VII
PERFORMANCE OF THE EVALUATION MODEL

| Model | Task | PCC | SORCC | RMSE |
|---|---|---|---|---|
| Baseline1 | LES | 0.976 | 0.960 | 0.170 |
| | VA | 0.966 | 0.979 | 0.219 |
| | BR | 0.956 | 0.963 | 0.288 |
| | TTT | 0.948 | 0.960 | 0.455 |
| | All | 0.902 | 0.868 | 0.303 |
| Baseline2 | LES | 0.993 | 0.960 | 0.079 |
| | VA | 0.988 | 0.979 | 0.184 |
| | BR | 0.988 | 0.914 | 0.189 |
| | TTT | 0.993 | 0.960 | 0.492 |
| | All | 0.929 | 0.863 | 0.282 |
| Baseline3 | LES | 0.984 | 0.960 | 0.175 |
| | VA | 0.973 | 0.960 | 0.181 |
| | BR | 0.987 | 0.914 | 0.274 |
| | TTT | 0.989 | 0.960 | 0.450 |
| | All | 0.922 | 0.862 | 0.292 |
| Proposed model (TPIFM) | LES | 0.997 | 0.979 | 0.089 |
| | VA | 0.996 | 0.979 | 0.109 |
| | BR | 0.994 | 0.963 | 0.156 |
| | TTT | 0.983 | 0.960 | 0.134 |
| | **All** | **0.985** | **0.986** | **0.124** |

Fig. 8 presents scatter plots of MOS and the predicted interactive quality to intuitively illustrate performance across different tasks. To evaluate task-dependent prediction accuracy, results are grouped by task type. As shown in Fig. 8, the proposed model consistently delivers accurate predictions across all tasks, with predicted scores closely matching subjective user ratings. This indicates that the proposed model effectively captures task-dependent characteristics and maintains robust predictive capability across varying interaction tasks. This is further supported by higher PCC and SROCC, as well as lower RMSE compared to the baseline models (see Table VII). Specifically, the proposed model achieves approximately 6.0% improvement in PCC and 13.6% in SROCC, along with a 51.6% reduction in RMSE compared to the best-performing baselines (Baseline2).

In contrast, the baseline models show varying performance across tasks. While they perform reasonably well for moderate JND tasks (e.g., VA and LES), their accuracy degrades noticeably in both low- and high-JNDs scenarios. Specifically, the baseline models exhibit a systematic bias in their predictions. They tend to over-predict PIF in low-JND tasks (e.g., BR), where users are highly sensitive to temporal deviations, and under-predict it in high-JND scenarios (e.g., TTT), where greater tolerance is observed. This behavior can be explained by their task-independent functional forms (Eqs. (7) and (8)). Without incorporating task-specific temporal sensitivity, these models fail to capture the stronger sensitivity to temporal deviations in low-JND scenarios and the greater tolerance to them in high-JND tasks. This observation is consistent with prior findings on collaborative stability under increased task complexity [26]. By contrast, the proposed model incorporates task-specific JND as a modulating factor in Eq. (2) and Eq. (5), enabling an adaptive mapping from network impairments to perceived fluency. This design allows the model to better capture task-dependent variations and improve prediction accuracy across different JND conditions.

### *B. Ablation Study*

Beyond the comparison with baseline models, we further conduct ablation studies to verify the contribution of each component in TPIFM. Five model configurations are constructed by removing or modifying individual components from the proposed formulation, including the full model, a variant without JND modulation (w/o JND), a variant without nonlinear JND-aware modulation (w/o nonlinear modulation), a variant without stalling modeling (w/o stalling model), and a variant without delay modeling (w/o delay model).

The impact of task-dependent perceptual sensitivity is first evaluated by removing the JND modulation. The w/o JND variant decreases the PCC and SROCC from 0.985 and 0.986 to 0.928 and 0.865, respectively, demonstrating that incorporating task-specific temporal sensitivity is essential for accurate PIF modeling. The contribution of nonlinear modulation is further evaluated by replacing the nonlinear JND-aware impairment functions with linear formulations while preserving the task-dependent JND information. The w/o nonlinear modulation variant also leads to performance degradation, achieving PCC and SROCC of 0.957 and 0.950, respectively. This suggests that nonlinear mapping better captures the relationship between temporal impairments and perceived fluency. For temporal impairments, removing either stalling or delay modeling decreases prediction accuracy. In particular, removing the stalling component causes the largest degradation, reducing PCC and SROCC to 0.492 and 0.534, respectively. This suggests that temporal interruptions play an essential role in shaping PIF in RCAR scenarios. Removing delay modeling results in a moderate performance decrease, indicating that delay also contributes to PIF perception but has a comparatively smaller impact. Overall, these results validate the necessity of each component in the proposed TPIFM.

TABLE VII
ABLATION STUDY OF THE EVALUATION MODEL

| Model settings | PCC | SORCC | RMSE |
|---|---|---|---|
| w/o JND | 0.928 | 0.865 | 0.256 |
| w/o nonlinear modulation | 0.957 | 0.950 | 0.259 |
| w/o stalling model | 0.492 | 0.534 | 0.862 |
| w/o delay model | 0.914 | 0.864 | 0.442 |
| Full model | **0.985** | **0.986** | **0.124** |

## VI. CONCLUSION

This study investigated how task-specific JNDs influence the perception of interaction fluency under delay and stalling impairments in RCAR scenarios. Through controlled subjective experiments, we demonstrated that network impairments negatively affect PIF, while task-dependent temporal sensitivity modulates the extent of such impact. Tasks with lower JNDs exhibited higher sensitivity to temporal impairments, whereas tasks with higher JNDs showed greater tolerance due to reduced temporal requirements. Based on these insights, a task-aware objective evaluation model (TPIFM) was proposed to assess PIF under delay and stalling conditions in RCAR scenarios. Experimental validation confirmed that TPIFM outperformed baseline models, improving PCC and SROCC by 6.0% and 13.6%, respectively, while reducing RMSE by over 50%.

Future work will further investigate the generalizability of TPIFM by extending the evaluation to more complex RCAR scenarios involving multi-task and interleaved interactions, different AR platforms, collaborative applications, more diverse task types, and broader participant populations, including professional users from representative RCAR domains. Furthermore, the proposed framework could be extended to multi-party collaborative telepresence scenarios involving richer social interactions in AR. In addition, integrating adaptive JND estimation into collaborative synchronization frameworks may further improve co-presence and collaboration efficiency in dynamic RCAR environments.

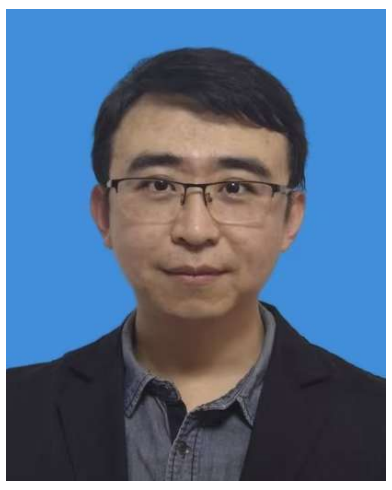

**Jiarun Song** (Member, IEEE**)** received the B.S. and Ph.D. degrees from Xidian University, Xi'an, China in 2009 and 2015, respectively. He is currently an Associate Professor with the State Key Laboratory of Integrated Services Networks, Xidian University. His research interests focus on QoE, video quality assessment, and multimedia communication.

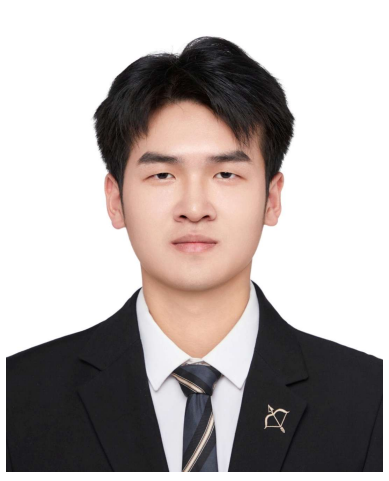

**Ninghao Wan** received the B.S. degree in Telecommunication Engineering from Xidian University, Xi'an, China, in 2024. He is currently a graduate student in the Xidian University. His research interests include Virtual/Augmented Reality, quality assessment, and multimedia communication.

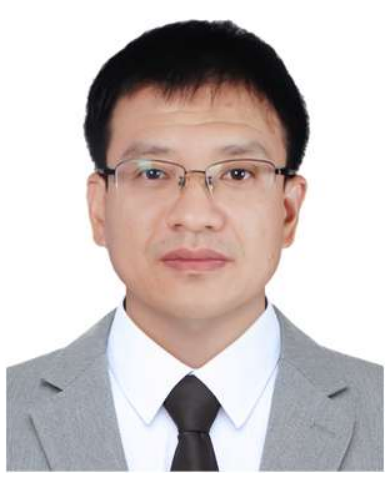

**Fuzheng Yang** (Member, IEEE) received the B.E. degree in Telecommunication Engineering, the M.E. degree and the Ph.D. in Communication and Information System from Xidian University, Xi'an, China, in 2000, 2003 and 2005, respectively. He became a lecturer and an Associate Professor in Xidian University in 2005 and 2006, respectively. He has been a professor of communications engineering with Xidian University since 2012. He is also an Adjunct Professor of School of Engineering in RMIT University. During 2006-2007, he served as a visiting scholar and postdoctoral researcher in Department of Electronic Engineering in Queen Mary, University of London. His research interests include video quality assessment, video coding and multimedia communication.

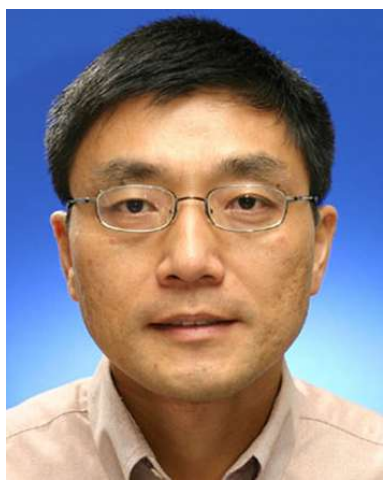

**Weisi Lin** (Fellow, IEEE) received the Ph.D. degree from the King's College, University of London, U.K. He is currently a Professor with the College of Computing and Data Science, Nanyang Technological University. His areas of expertise include image processing, perceptual signal modeling, video compression, and multimedia communication, in which he has published over 200 journal articles, over 230 conference papers, filed seven patents, and authored two books. He has been an invited/panelist/keynote/tutorial speaker at over 20 international conferences. He is a fellow of IET and an Honorary Fellow of Singapore Institute of Engineering Technologists. He has been the Technical Program Chair of IEEE ICME 2013, PCM 2012, QoMEX 2014, and IEEE VCIP 2017. He has been an Associate Editor of IEEE TRANSACTIONS ON IMAGE PROCESSING, IEEE TRANSACTIONS ON CIRCUITS AND SYSTEMS FOR VIDEO TECHNOLOGY, IEEE TRANSACTIONS ON MULTIMEDIA, and IEEE SIGNAL PROCESSING LETTERS. He was a Distinguished Lecturer of Asia-Pacific Signal and Information Processing Association (APSIPA) from 2012 to 2013 and the IEEE Circuits and Systems Society from 2016 to 2017.